\documentclass[osajnl,twocolumn,showpacs,superscriptaddress,10pt,letterpaper]{revtex4-1}
\usepackage{amsmath,amssymb,graphicx}
\usepackage{epstopdf}
\begin{document}

\title{Stokes--anti-Stokes Correlations in Raman Scattering from Diamond Membranes}

\author{Mark Kasperczyk}
\affiliation{Photonics Laboratory, ETH Z\"{u}rich, 8093 Z\"{u}rich, Switzerland}

\author{Ado Jorio}
\affiliation{Departamento de F\'{\i}sica, Universidade Federal de Minas Gerais, Belo Horizonte, MG 30123-970, Brazil}

\author{Elke Neu}
\author{Patrick Maletinsky}
\affiliation{Department of Physics, University of Basel, Klingelbergstrasse 82, CH-4056 Basel, Switzerland}

\author{Lukas Novotny}
\email{Corresponding author: lnovotny@ethz.ch}
\affiliation{Photonics Laboratory, ETH Z\"{u}rich, 8093 Z\"{u}rich, Switzerland}

\begin{abstract}
We investigate the arrival statistics of Stokes (S) and anti-Stokes (aS) Raman photons generated in diamond membranes.
Strong quantum correlations between the S and aS signals are observed, which implies that the two processes share the same phonon, that is, the phonon excited by the S process is consumed in the aS process. We show that the intensity cross-correlation $g_{\rm S,aS}^{(2)}(0)$, which describes the simultaneous detection of Stokes and anti-Stokes photons,
decreases steadily with laser power as $1/{\rm P_L}$. Contrary to many other material systems, diamond exhibits a maximum $g_{\rm S,aS}^{(2)}(0)$ at very low pump powers, implying that the Stokes-induced aS photons outnumber the thermally generated aS photons. On the other hand, the coincidence rate shows a quadratic plus cubic power dependence, which indicates a departure from the Stokes-induced anti-Stokes process.
\end{abstract}

\maketitle
\begin{figure}[b!]
\includegraphics[width=0.30\textwidth]{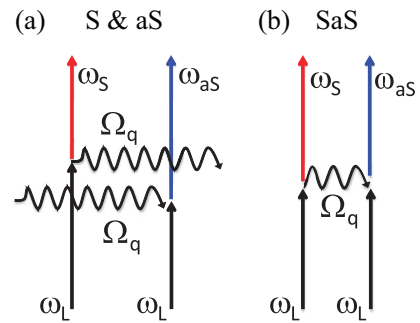}%
\vspace{-0.5em}
\caption{Schematic representation of Stokes (S) and anti-Stokes (aS) Raman scattering. (a) Uncorrelated S and aS processes (S\&aS). The phonons responsible for the aS process are generated thermally. (b) In the absence of thermal phonons, anti-Stokes photons are generated by the phonons created through the Stokes process. Stokes photons and Stokes-induced anti-Stokes (SaS) photons become correlated.\label{fig1}}
\end{figure}

Raman scattering, conventionally used as a method for probing the vibrational modes of a system, can be used to create correlated Stokes--anti-Stokes photon pairs \cite{Wal-12,Suss-13,Wal-112,Wal-11,Lukin-03,Wal-10,Kimble-03,Most-81,Duan-02,Klysh-77,Berto-84}. In the uncorrelated regime, Raman scattering is spontaneous and, without laser heating, both Stokes and anti-Stokes intensities are linear with excitation laser power (see Fig.~\ref{fig1}a). However, if the phonon energies are high enough that the thermal phonon occupation is low, the spontaneous aS process is rare and correlations between Stokes and anti-Stokes photon generation set in (see in Fig.~\ref{fig1}b). In this regime the aS intensity depends on the squared excitation laser power, since one laser photon writes the phonon, and another laser photon reads it. Recent work has shown both theoretically and experimentally that the Stokes-generated phonon acts as a quantum memory, where the Stokes and anti-Stokes signals act as write and read commands \cite{Wal-12,Suss-13,Wal-112,Wal-11,Lukin-03,Wal-10,Kimble-03,Most-81,Duan-02,Klysh-77,Berto-84}. In parallel, research in photon pairs produced through four-wave mixing (FWM) in optical fibers has shown extremely high nonclassical correlations \cite{
Wang-052,Wang-01}, analogous to studies in spontaneous parametric downconversion (SPDC)~\cite{Akiba-09}. \\[-1.5ex]

In this paper, we report the generation of highly nonclassical photon superbunching in diamond at low excitation powers and analyze Stokes--anti-Stokes photon correlations as a function of pump power. Our data reveal the range of conditions under which Stokes-induced anti-Stokes scattering (SaS) can be used to generate correlated photons in diamond. This information is useful for designing efficient phonon-based quantum memories and heralded single-photon sources for quantum communication. Contrary to four-wave mixing measurements in optical fibers \cite{
Wang-052} and spontaneous parametric down conversion in nonlinear crystals \cite{Akiba-09}, we observe a saturation of Stokes--anti-Stokes correlations at very low intensities.  \\[-1.5ex]

\begin{figure}[!b]
\includegraphics[width=0.35\textwidth]{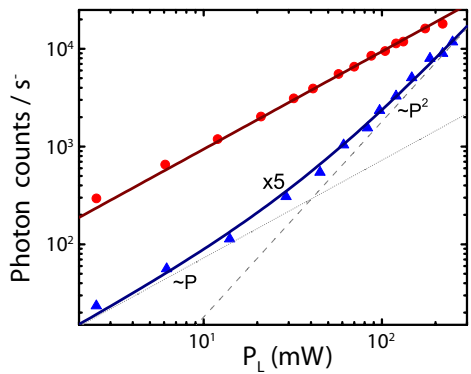}%
\vspace{-0.5em}
\caption{Dependence of Stokes signal (red circles) and anti-Stokes signal (blue triangles) on average excitation power measured in the forward direction. The data points are the area of Gaussian fits to the spectral lines, normalized by the integration time. The Stokes signal is linear with laser power, whereas the anti-Stokes signal shows a linear power dependence for low laser powers and quadratic dependence for high laser powers, suggesting that the SaS process becomes much stronger than spontaneous aS scattering. The aS data and fit have been multiplied by a factor of 5 to show it on the same scale as the S data. The two gray lines are the linear (dotted) and quadratic (dashed) components of the aS fit.\label{fig2}}
\end{figure}

In our experiments, we measure Stokes and anti-Stokes photons in diamond as a function of laser power ${\rm P_L}$. The excitation wavelength is $\lambda = 785\,$nm and the Stokes and anti-Stokes photons appear at the wavelengths $\lambda_{\rm S} = 880\,$nm and $\lambda_{\rm aS} = 710\,$nm, respectively, as defined by the phonon frequency of $1332\,$cm$^{-1}$ in diamond. The duration of the excitation laser pulses is $\tau=130\,$fs, with a repetition rate of $\Delta f = 76\,$MHz. The sample is a freestanding $50\,\mu$m thick diamond membrane. As shown in Fig.~\ref{fig2}, the Stokes signal shows a linear power dependence, whereas the anti-Stokes signal exhibits a quadratic power dependence at high intensities, as expected for the SaS process illustrated in Fig.~\ref{fig1}b. At low powers, on the other hand, the spontaneously generated aS process is stronger than the SaS process, and hence the aS signal shows a linear power dependence.  \\[-1.5ex]

To understand the interplay between the spontaneous and the Stokes-induced aS processes, we measure the second-order intensity cross-correlation $g_{\rm S,aS}^{(2)}(0)$ as a function of average laser power. $g_{\rm S,aS}^{(2)}(0)$ corresponds to the total number of Stokes and anti-Stokes coincidences measured at zero time delay, normalized by the number of accidental coincidences. Classically, $g_{\rm S,aS}^{(2)}(0)$ is bound by the products of the autocorrelations through the inequality $g^{(2)}_{\rm S,aS}(0) \leq \sqrt{g^{(2)}_{\rm S,S}(0)\, g^{(2)}_{\rm aS,aS}(0)}$ \cite{Wal-12,Loud-10}. In turn, the autocorrelations are classically bound by $g^{(2)}_{\rm S,S}(0), g^{(2)}_{\rm aS,aS}(0) \leq 2$, where the equality holds for a thermal source \cite{Wal-12,Loud-10}. For a large range of intensities, the SaS process dominates and the value of $g^{(2)}_{\rm S,aS}(0)$ strongly violates the inequality since the photon arrival statistics cannot be described classically.  \\[-1.5ex]

\begin{figure}[!t]
\includegraphics[width=0.45\textwidth]{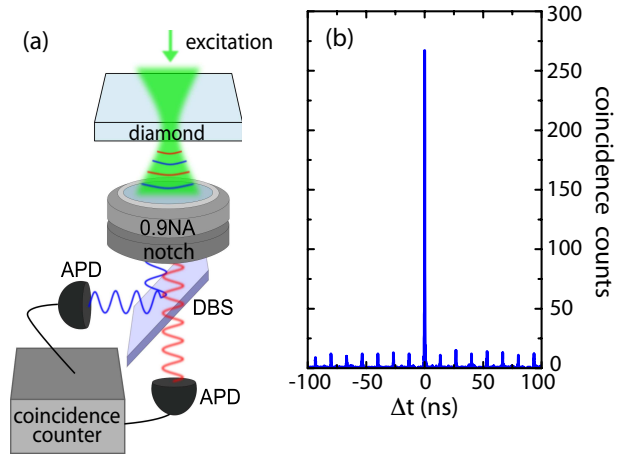}%
\vspace{-0.5em}
\caption{(a) Experimental setup. A $0.5\,$NA objective (not shown) focuses $785\,$nm light (shown in green) from a Ti:Sapph laser into a $50\,\mu$m thick diamond sample. The scattered light is collected with a $0.9\,$NA air objective. The light can either be sent to a spectrometer (not shown), or to a pair of avalanche photodiodes (APD). A dichroic beamsplitter (DBS) separates the Stokes (red) and anti-Stokes (blue) signals, sending them to separate APDs. 
(b) A typical coincidence measurement. 
Count rates for this experiment were $4.2\,$kHz for Stokes and $200\,$Hz for anti-Stokes. The incident power was $8.6\,$mW\label{fig3}}
\end{figure}

\begin{figure}[!b]
\includegraphics[width=0.42\textwidth]{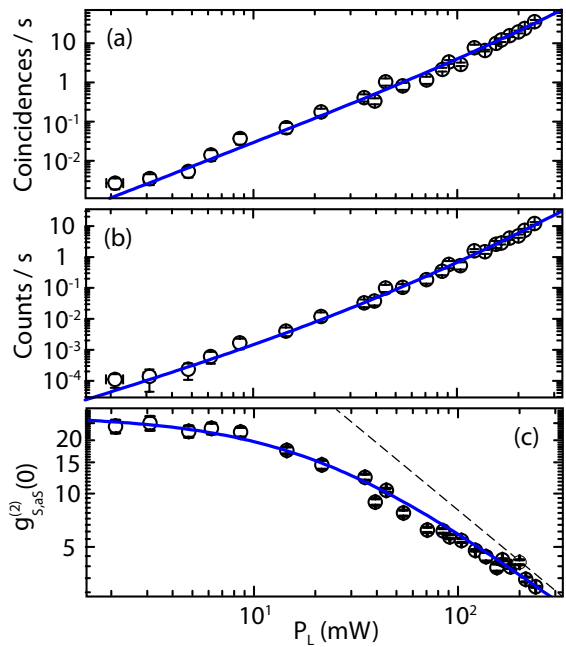}%
\vspace{-0.5em}
\caption{Correlation of Stokes and anti-Stokes photons as a function of laser power. (a) Coincidence rate at $\Delta t = 0$. 
The fitting curve is quadratic for low P and cubic for high P. (b) Coincidence rate at $\Delta t \neq 0$. 
(c) Second-order correlation for zero time delay $g_{\rm S,aS}^{(2)}(0)$. 
The dashed line indicates the asymptotic $1/{\rm P_L}$ behavior. At low powers, $g_{\rm S,aS}^{(2)}(0)$ stops increasing because the yield of Stokes-induced anti-Stokes photons becomes comparable to the yield of thermally induced anti-Stokes photons.\label{fig4}}
\end{figure}

Our experimental setup for measuring correlations between Stokes and anti-Stokes photons is illustrated in Fig.~\ref{fig3}a. A representative coincidence measurement is shown in Fig.~\ref{fig3}b. The coincidence rate is calculated by dividing the coincidence counts at time delay $\Delta t = 0$ by the measurement time. As shown in Fig.~\ref{fig4}a, the power dependence of the coincidence rate is mostly quadratic, but for very high powers it becomes necessary to include a cubic term in the fit. Using the recorded coincidence counts we can evaluate the second-order cross-correlation function $g_{\rm S,aS}^{(2)}(0)$ by dividing the peak at $\Delta t = 0$ by the average of the peaks at $\Delta t \neq 0$. $g_{\rm S,aS}^{(2)}(0)$ initially increases as the power is decreased and then eventually reaches a maximum value (see Fig.~\ref{fig4}c).  \\[-1.5ex]

The coincidence rate and the cross-correlation $g_{\rm S,aS}^{(2)}(0)$ can both be written in terms of probabilities of generating Stokes and anti-Stokes photons. We assume that the number of Stokes photons generated per pulse is small enough that the probability of generating a Stokes photon is always linear with laser power, which ensures that we are not in the regime of stimulated Raman scattering (SRS). We further assume that the number of thermal phonons in diamond is negligible compared to the number of Stokes-induced phonons. This assertion is justified by the fact that the thermal population of phonons in diamond is small at room temperature because of diamond's large phonon energy ($1332\,$cm$^{-1}$, or $40\,$THz). Therefore, if we detect an anti-Stokes photon, we know it was produced by the SaS process. This assumption should fail at very low powers, when the number of Stokes-induced phonons becomes comparable to the number of thermal phonons. Given these considerations, we can explain the power dependence of the number of coincidences per second through the probability of measuring a Stokes--anti-Stokes pair $P(\rm S,aS)$. An application of the product rule allows this to be rewritten as
\begin{equation}
P({\rm S,aS})\,=\,P({\rm aS \vert S})\,P({\rm S}) \,=\, P({\rm S \vert aS})\, P({\rm aS})\, ,
\end{equation}
where $P(\rm aS \vert S)$ denotes the conditional probability of detecting an anti-Stokes photon given that a Stokes photon has already been detected, and similarly for $P(\rm S \vert aS)$ \cite{Wal-12}. $P(\rm S)$ and $P(\rm aS)$ are the unconditional probabilities for detecting Stokes and anti-Stokes photons. Because of the absence of thermal phonons, $P(\rm S \vert aS)$ is given by the collection and detection efficiency $\eta_S$ at the Stokes frequency. In other words, detection of an anti-Stokes photon requires that a Stokes photon was created, though it may be undetected. The power dependence of the coincidences is therefore ruled by the power dependence of the anti-Stokes signal, that is, $P({\rm S,aS}) = \eta_S P({\rm aS})$. In the absence of thermal phonons,  the anti-Stokes signal scales quadratically with  laser power ${\rm P_L}$ (see Fig.~\ref{fig2}) and hence  $P({\rm S,aS}) \propto {\rm P_L}^2$, which is confirmed by our coincidence measurements shown in Fig.~\ref{fig4}a.  Note, however, that spontaneously generated coincidences also exhibit a quadratic power dependence. In this case, Stokes and anti-Stokes processes are independent, that is,  $P({\rm S \vert aS}) = P({\rm S})$, and hence  $P({\rm S,aS})= P({\rm S})\, P({\rm aS}) \propto {\rm P_L}^2$. Different to the SaS process, spontaneous coincidences are independent of time delay $\Delta t$ and therefore their contribution is always smaller than the height of the off-center peaks in Fig.~\ref{fig3}b.  \\[-1.5ex]

At very high pump powers (${\rm P_L} > 150\,$mW), we observe the onset of a cubic term in the coincidence rate. At these high powers, we need to account for stimulated processes, such as coherent Raman scattering (CRS). Alternatively, for strong interactions, the system can undergo multiple phonon-photon swapping cycles as the beam propagates through the sample {\cite{Gall-14}}. Furthermore, once the anti-Stokes signal becomes comparable to the Stokes signal, the aS signal will increase more slowly with power, since in this regime most of the available Stokes-induced phonons are converted into anti-Stokes photons. Any of these contributions can lead to the observed deviations from the expected ${\rm P_L}^2$ dependence of the coincidence rate at high powers.  \\[-1.5ex]

We can use a similar analysis to understand the power dependence of the second-order cross-correlation \cite{Wal-12}
\begin{equation}
g_{\rm S,aS}^{(2)}(0) \,=\, \frac{P({\rm S,aS})}{P({\rm S}) \,P({\rm aS})} \,=\, \frac{P({\rm S \vert aS})}{P({\rm S})}.
\end{equation}
As noted above, $P(\rm S \vert aS) \approx \eta_S$, which leads to $g_{\rm S,aS}^{(2)}(0) = \eta_S\,/\,P(\rm S) = \eta_S\,/\,({\it k}_{\rm S} {\it P})$, where $k_{\rm S}$ is a constant describing the strength of the Stokes process as well as the collection and detection efficiencies. The correlation is therefore expected to increase with decreasing laser power as $1/{\rm P_L}$. At very high powers, we can no longer exclude the generation of multiple photon pairs per pulse and hence the value of $g^{(2)}_{\rm S,aS}(0)$ becomes dependent on the photon pair statistics. This can be understood as a transition from a single quantum state at low powers to an ensemble state which can be described classically at high power. At very low powers, the SaS process is no longer much stronger than the thermally generated anti-Stokes, which leads to a maximum attainable $g_{\rm S,aS}^{(2)}(0)$. As the laser power is further reduced, we expect that eventually the value of $g_{\rm S,aS}^{(2)}(0)$ will decrease, since the uncorrelated, spontaneously generated Stokes and anti-Stokes photons will dominate the signals. However, the integration time for such a measurement is prohibitively long. In the absence of thermally generated anti-Stokes scattering we would expect that $g_{\rm S,aS}^{(2)}(0)$ will increase without limit as power is reduced, similar to SPDC~\cite{Akiba-09}.  \\[-1.5ex]

Even at low pump powers, a significant number of phonons is produced per pulse ({$\sim$}4 phonons per pulse at $5\,$mW) \cite{Grim-75}. Because of momentum conservation, only a small fraction of these phonons are actually available for generating anti-Stokes photons \cite{Most-81,Soren-09}. An accurate estimate of when the spontaneous aS and SaS processes are equal therefore requires a careful consideration of the phase-matching conditions, as well as experimental factors such as collection volume and NA.  \\[-1.5ex]

It may seem unintuitive that $g_{\rm S,aS}^{(2)}(0)$ falls with increasing laser power even though the coincidences per second rise with power. The reason is that while the coincidences at $\Delta t = 0$ (center peak in Fig.~\ref{fig3}b) rise as ${\rm P_L}^2$ or ${\rm P_L}^3$ (depending on pump power), the coincidences for $\Delta t \neq 0$ (off-center peaks in Fig.~\ref{fig3}b) rise even faster. These are the correlations between Stokes and anti-Stokes photons excited by different laser pulses. Consider, for example, the probability of measuring a Stokes photon from one pulse and an anti-Stokes photon from the pulse immediately following. This corresponds to the peak at $\Delta t = 13\,$ns in figure 3(b). It requires that a Stokes photon is generated in one pulse, followed by a Stokes and anti-Stokes photon generated in the subsequent pulse, that is, although we detect only one Stokes photon from the first pulse and one anti-Stokes photon from the next pulse, a Stokes photon must still be generated in the second pulse for the anti-Stokes photon to be created. In total, the peak at $\Delta t = 13\,$ns includes the generation of a second Stokes photon, which implies that its power dependence is one order higher than the peak at $\Delta t = 0\,$ns. This argument holds for any Stokes--anti-Stokes pair that is not produced within the same pulse. The conclusion is that the peaks at $\Delta t \neq 0$ rise faster than the peaks at $\Delta t = 0$ by a rate proportional to $P$. When we calculate $g_{\rm S,aS}^{(2)}(0)$ by dividing the peak $\Delta t = 0$ by the average of the peaks at $\Delta t \neq 0$, the ratio will therefore fall as $P^{-1}$, which agrees with our measurements shown in Fig.~\ref{fig4}c.  Furthermore, using a thinner sample will reduce the peaks at $\Delta t \neq 0$ more than the peaks at $\Delta t = 0$. It is therefore beneficial to use membranes to achieve values of $g_{\rm S,aS}^{(2)}(0)$ far greater than 2.  \\[-1.5ex]

In conclusion, by studying the correlations between Stokes and anti-Stokes photons as a function of laser pump power, we have shown that the second-order cross-correlation $g^{(2)}_{\rm S,aS}(0)$ can be varied over a large range of values. We have discussed distinct regimes for producing Stokes--anti-Stokes photon pairs: 1) uncorrelated photons from thermal phonons and 2) correlated photons from the SaS process. The relative strengths of these processes depends on experimental parameters (particularly pump power), as well as material properties. For quantum computing, the optimal material should be in the SaS regime for a large range of pump powers, which can be achieved, for example, by exploiting material resonances. Future work will therefore focus on developing engineered samples~\cite{Elke-14} with distinct resonances, such as optical cavities~\cite{hausmann12}.  \\[-1.5ex]

The authors acknowledge fruitful and stimulating discussions with P. Bharadwaj. A.J. acknowledges CAPES and ETH for financing his visit to ETH Z\"{u}rich. L.N. thanks the Swiss National Science Foundation (SNF) for financial support through grant  200021-146358.\\


\end{document}